\documentclass[english,reprint]{revtex4-1}
\usepackage[T1]{fontenc}
\usepackage[latin9]{inputenc}
\usepackage{graphicx}
\usepackage{esint}

\makeatletter

\providecommand{\tabularnewline}{\\}

\usepackage{url}

\renewcommand\[{\begin{equation}}
\renewcommand\]{\end{equation}}

\usepackage{hyperref}
\hypersetup{
    colorlinks,
    citecolor=black,
    filecolor=black,
    linkcolor=black,
    urlcolor=black
}

\makeatother

\usepackage{babel}
\begin{document}

\author{Zack Lasner}
\email{zack.lasner@yale.edu}

\author{D. DeMille}

\affiliation{Yale University, New Haven, CT}

\begin{abstract}
In atomic and molecular phase measurements using laser-induced fluorescence
detection, optical cycling can enhance the effective photon detection
efficiency and hence improve sensitivity. We show that detecting many
photons per atom or molecule, while necessary, is not a sufficient
condition to approach the quantum projection limit for detection of
the phase in a two-level system. In particular, detecting the maximum
number of photons from an imperfectly closed optical cycle reduces
the signal-to-noise ratio (SNR) by a factor of $\sqrt{2}$, compared
to the ideal case in which leakage from the optical cycle is sufficiently
small. We derive a general result for the SNR in a system in terms
of the photon detection efficiency, probability for leakage out of
the optical cycle per scattered photon, and the product of the average
photon scattering rate and total scattering time per atom or molecule.
\end{abstract}

\title{Statistical sensitivity of phase measurements via laser-induced fluorescence
with optical cycling detection}

\maketitle
Atoms and molecules are powerful platforms to probe phenomena at quantum-projection-limited
precision. In many atomic and molecular experiments, a quantum state
is read out by laser-induced fluorescence (LIF), in which population
is driven to a short-lived state and the resulting fluorescence photons
are detected. Due to geometric constraints on optical collection and
technological limitations of photodetectors, the majority of emitted
photons are typically undetected, reducing the experimental signal.
Optical cycling transitions can be exploited to overcome these limitations,
by scattering many photons per particle. In the limit that many photons
from each particle are detected, the signal-to-noise ratio (SNR) may
be limited by the quantum projection (QP) noise (often referred to
as atom or molecule shot noise). LIF detection with photon cycling
is commonly used in ultra-precise atomic clock \cite{Wynands2005,Zelevinsky2008}
and atom interferometer \cite{Cronin2009} experiments to approach
the QP limit.

Molecules possess additional features, beyond those in atoms, that
make them favorable probes of fundamental symmetry violation \cite{ACMECollaboration2014,Collaboration2018,Hudson2011,Devlin2015,Hunter2012,Kozyryev2017}
and fundamental constant variation \cite{Borkowski2018,Beloy2011,DeMille2008,Zelevinsky2008,Shelkovnikov2008,Kozyryev2018},
as well as promising platforms for quantum information and simulation
\cite{DeMille2002,Liu2018,Micheli2006,Sundar2018,Wall2015}. Many
molecular experiments that have been proposed, or which are now being
actively pursued, will rely on optical cycling to enhance measurement
sensitivity while using LIF detection \cite{Collaboration2018,Hunter2012,Kozyryev2018,Kozyryev2017,ACMECollaboration2014,Devlin2015}.
Due to the absence of selection rules governing vibrational decays,
fully closed molecular optical cycling transitions cannot be obtained:
each photon emission is associated with a non-zero probability of
decaying to a ``dark'' state that is no longer driven to an excited
state by any lasers. However, for some molecules many photons can
be scattered using a single excitation laser, and up to $\sim10^{6}$
photons have been scattered using multiple repumping lasers to return
population from vibrationally excited states into the optical cycle
\cite{DiRosa2004,Shuman2009}. This has enabled, for example, laser
cooling and magneto-optical trapping of molecules \cite{Shuman2010,Barry2014,Hummon2013,Collopy2018,Zhelyazkova2014,Truppe2017,Chae2017,Anderegg2017}.
Furthermore, some precision measurements rely on atoms in which no
simply closed optical cycle exists \cite{Regan2002,Parker2015}; our
discussion here will be equally applicable to such species.

These considerations motivate a careful study of LIF detection for
precision measurement under the constraint of imperfectly closed optical
cycling. Some consequences of loss during the cycling process have
been considered in \cite{Rocco2014}. However, the effect of the statistical
nature of the cycling process on the optimal noise performance has
not been previously explored. In particular, the number of photons
scattered before a particle (an atom or molecule) decays to an unaddressed
dark state, and therefore ceases to fluoresce, is governed by a statistical
distribution rather than a fixed finite number. We show that due to
the width of this distribution, a naive cycling scheme reduces the
SNR to below the QP limit. In particular, we find that in addition
to the intuitive requirement that many photons from every particle
are detected, to approach the QP limit it is also necessary that the
probability of each particle exiting the cycling transition (via decay
to a dark state outside the cycle) is negligible during detection.
If this second condition is not satisfied, so that each particle scatters
enough photons that it is very likely to have been optically pumped
into a dark state, then the SNR is decreased by a factor of $\sqrt{2}$
below the QP limit.

Consider an ensemble of $N$ particles in an effective two-level system,
in a state of the form
\[
|\psi\rangle=(e^{-i\phi}|\uparrow\rangle+e^{i\phi}|\downarrow\rangle)/\sqrt{2}.
\]
The relative phase $\phi$ is the quantity of interest in this discussion.
It can be measured, for example, by projecting the wavefunction onto
an orthonormal basis $\{|X\rangle\propto|\uparrow\rangle+|\downarrow\rangle,\,|Y\rangle\propto|\uparrow\rangle-|\downarrow\rangle\}$
such that $|\langle X|\psi\rangle|^{2}=\cos^{2}(\phi)$ and $|\langle Y|\psi\rangle|^{2}=\sin^{2}(\phi)$.
In the LIF technique, this can be achieved by driving state-selective
transitions, each addressing either $|X\rangle$ or $|Y\rangle$,
through an excited state that subsequently decays to a ground state
and emits a fluorescence photon. This light is detected, and the resulting
total signals, $S_{X}$ and $S_{Y}$, are associated with each state.
(This protocol is equivalent to the more standard Ramsey method, in
which each spin is reoriented for detection by a spin-flip pulse and
the population of spin-up and spin-down particles is measured \cite{Ramsey1950}.)
The measured value of the phase, $\tilde{\phi},$ is computed from
the observed values of $S_{X}$ and $S_{Y}$. In the absence of optical
cycling, the statistical uncertainty of the phase measurement is $\sigma_{\tilde{\phi}}=\frac{1}{2\sqrt{N\epsilon}}$,
where $\epsilon$ is the photon detection efficiency and $0<\epsilon\leq1$.
Note that $N\epsilon$ is the average number of detected photons;
hence, this result is often referred to as the ``photon shot noise
limit.'' In the ideal case of $\epsilon=1$, the QP limit (a.k.a.
the atom or molecule shot noise limit) limit $\sigma_{\tilde{\phi}}=\frac{1}{2\sqrt{N}}$
is obtained. This scaling is derived as a limiting case of our general
treatment below, where the effects of optical cycling are also considered.

We suppose that the phase is projected onto the $\{|X\rangle,\,|Y\rangle\}$
basis independently for each particle. Repeated over the ensemble
of particles, the total number of particles $N_{X}$ projected along
$|X\rangle$ is drawn from a binomial distribution, $N_{X}\sim B(N,\,\cos^{2}\phi)$,
where $x\sim f(\alpha_{1},\cdots,\alpha_{k})$ denotes that the random
variable $x$ is drawn from the probability distribution $f$ parametrized
by $\alpha_{1},\cdots,\alpha_{k}$, and $B(\nu,\,\rho)$ is the binomial
distribution for the total number of successes in a sequence of $\nu$
independent trials that each have a probability $\rho$ of success.
Therefore, $\overline{N_{X}}=N\,\cos^{2}\phi$ and $\sigma_{N_{X}}^{2}=N\,\cos^{2}\phi\sin^{2}\phi$,
where $\bar{x}$ is the expectation value of a random variable $x$
and $\sigma_{x}$ is its standard deviation over many repetitions
of an experiment. We define the number of photons scattered from the
$i$-th particle to be $n_{i}$, where a ``photon scatter'' denotes
laser excitation followed by emission of one spontaneous decay photon,
and define $\overline{n_{i}}=\bar{n}$ (the average number of photons
scattered per particle) and $\sigma_{n_{i}}=\sigma_{n}$. Note that
these quantities are assumed to be the same for all particles (i.e.,
independent of $i$). The probability of detecting any given photon
(including both imperfect optical collection and detector quantum
efficiency) is $\epsilon$, such that each photon is randomly either
detected or not detected. We define $d_{ij}$ to be a binary variable
indexing whether the $j$-th photon scattered from the $i$-th particle
is detected. Therefore, $d_{ij}\sim B(1,\,\epsilon)$, and it follows
that $\overline{d_{ij}}=\epsilon$ and $\sigma_{d_{ij}}^{2}=\epsilon(1-\epsilon)$.

We define the signal of the measurement of a particular quadrature
$|X\rangle$ or $|Y\rangle$ from the ensemble, when projecting onto
that quadrature, to be the total number of photons detected. For example,
the signal $S_{X}$ from particles projected along $|X\rangle$ is
\begin{equation}
S_{X}=\sum_{i=1}^{N_{X}}\sum_{j=1}^{n_{i}}d_{ij}.\label{eq:Sx definition}
\end{equation}
$\noindent$Explicitly, among $N$ total particles, $N_{X}$ are projected
by the excitation light onto the $|X\rangle$ state and the rest are
projected onto $|Y\rangle$. The $i$-th particle projected onto $|X\rangle$
scatters a total of $n_{i}$ photons, and we count each photon that
is detected (in which case $d_{ij}=1)$. The right-hand side of Eq.
\ref{eq:Sx definition} depends on $\phi$ implicitly through $N_{X}$,
and we use this dependence to compute $\tilde{\phi}$, the measured
value of $\phi$. Because $N_{X},\,n_{i},$ and $d_{ij}$ are all
statistical quantities, the extracted value $\tilde{\phi}$ has a
statistical uncertainty. The QP limit is achieved when the only contribution
to uncertainty arises from $N_{X}$ due to projection onto the $\{|X\rangle,|Y\rangle\}$
basis.

We can compute $\overline{S_{X}}$ by repeated application of Wald's
lemma (\cite{Bruss1991,Wald2013}), $\overline{\sum_{i=1}^{m}x}=\bar{m}\bar{x}$.
This results in

\begin{equation}
\overline{S_{X}}=N\cos^{2}\phi\,\bar{n}\epsilon.\label{eq:ESx}
\end{equation}
$\noindent$That is, the expected signal from projecting onto the
$|X\rangle$ state is (as could be anticipated) simply the product
of the average number of particles in $|X\rangle$, $N\cos^{2}\phi$,
the number of photons scattered per particle, $\bar{n}$, and the
probability of detecting each photon, $\epsilon$.

We compute the variance in $S_{X}$ by repeated use of the law of
total variance \cite{Blitzstein}, $\sigma_{a}^{2}=\overline{\sigma_{a|b}^{2}}+\sigma_{\overline{a|b}}^{2}$,
where $\overline{a|b}$ denotes the mean of $a$ conditional on a
fixed value of $b$ and, analogously, $\sigma_{a|b}^{2}$ denotes
the variance of $a$ conditional on a fixed value of $b$. This gives

\[
\sigma_{S_{X}}^{2}=N\cos^{2}\phi\,\bar{n}\epsilon^{2}\left(\frac{1}{\epsilon}+\frac{\sigma_{n}^{2}}{\bar{n}}-1+\bar{n}\sin^{2}\phi\right).
\]
$\noindent$The results for $S_{Y}$ are identical, with the substitution
$\cos^{2}\phi\leftrightarrow\sin^{2}\phi$. Many atomic clocks \cite{Weyers2001,Jefferts2002,Kurosu2004,Levi2004,Szymaniec2005a}
and some molecular precision measurement experiments \cite{ACMECollaboration2014,Devlin2015}
measure both $S_{X}$ and $S_{Y}$, while others detect only a single
state \cite{Collaboration2018,Hudson2011,Regan2002,Parker2015}. In
what follows, we assume that both states are probed. The case of detecting
only one state, with some means of normalizing for variations in $N\bar{n}\epsilon$,
can be worked out using similar considerations.

In the regime $\phi=\pm\frac{\pi}{4}+\delta\phi$, where $\delta\phi\ll1$,
sensitivity to small changes in phase, $\delta\phi$, is maximized.
In this case, we define the measured phase deviation $\delta\tilde{\phi}$
by $\tilde{\phi}=\pm\frac{\pi}{4}+\delta\tilde{\phi}$. This is related
to measured quantities via the asymmetry $\mathcal{A}=\frac{S_{X}-S_{Y}}{S_{X}+S_{Y}}=\mp\sin(2\delta\tilde{\phi})\approx\mp2\delta\tilde{\phi}$.
When $N\gg1$, the average value of $\tilde{\phi}$ computed in this
way is equal to the phase $\phi$ of the two-level system.

The uncertainty in the asymmetry, $\sigma_{\mathcal{A}}\approx\frac{1}{N}\sqrt{\sigma_{S_{X}}^{2}+\sigma_{S_{Y}}^{2}-2\sigma_{S_{X},S_{Y}}^{2}}$,
can be computed to leading order in $\delta\phi$ from $\sigma_{S_{X}}$,
$\sigma_{S_{Y}}$, and the covariance $\sigma_{S_{X},S_{Y}}^{2}=\overline{S_{X}S_{Y}}-\overline{S_{X}}\,\overline{S_{Y}}$
using standard error propagation \cite{Bevington1969}. We relate
$\sigma_{\mathcal{A}}$ to the uncertainty in the measured phase by
$\sigma_{\mathcal{A}}=2\sigma_{\tilde{\phi}}$. This relationship
defines the statistical uncertainty in $\tilde{\phi}$, the measured
value of $\phi$, for the protocol described here. The covariance,
$\sigma_{S_{X},S_{Y}}^{2}=-\frac{N}{4}\bar{n}^{2}\epsilon^{2}$, can
be calculated directly using the same methods already described. This
result can be understood as follows: the photon scattering and detection
processes for particles projected onto $|X\rangle$ and $|Y\rangle$
are independent, so the covariance between signals $S_{X}$ and $S_{Y}$
only arises from quantum projection. In the simplest case of perfectly
efficient, noise-free detection and photon scattering, e.g., $\epsilon=1$,
$\bar{n}=1$, and $\sigma_{n}=0$, the quantum projection noise leads
to signal variances $\sigma_{S_{X}}^{2}=\sigma_{S_{Y}}^{2}=\frac{N}{4}$.
The covariance is negative because a larger number of particles projected
onto $|X\rangle$ is associated with a smaller number of particles
projected onto $|Y\rangle$. The additional factor of $\bar{n}^{2}\epsilon^{2}$
for the general case accounts for the fact that both signals $S_{X}$
and $S_{Y}$ are scaled by $\bar{n}\epsilon$ when $\bar{n}$ photons
are scattered per particle and a proportion $\epsilon$ of those photons
are detected on average.

The uncertainty in the measured phase, computed using the procedure
just described, has the form $\sigma_{\tilde{\phi}}=\frac{1}{2\sqrt{N}}\sqrt{F}$,
where we have defined the ``excess noise factor'' $F$ given in
this phase regime by

\[
F=1+\frac{1}{\bar{n}}\left(\frac{1}{\epsilon}-1\right)+\frac{\sigma_{n}^{2}}{\bar{n}^{2}}.
\]

It is instructive to evaluate this expression in some simple limiting
cases. For example, consider the case when exactly one photon is scattered
per particle so that $\bar{n}=1$ and $\sigma_{n}=0$. (This is typical
for experiments with molecules, where optical excitation essentially
always leads to decay into a dark state.) In this case, $F=\frac{1}{\epsilon}$
and the uncertainty in the phase measurement is $\sigma_{\tilde{\phi}}=\frac{1}{2\sqrt{N\epsilon}}$,
as stated previously. Alternatively, as $\bar{n}\rightarrow\infty$,
$F\rightarrow1+\left(\frac{\sigma_{n}}{\bar{n}}\right)^{2}$. This
is in exact analogy with the excess noise of a photodetector whose
average gain is $\bar{n}$ and whose variance in gain is $\sigma_{n}^{2}$
\cite{Knoll2010}. By inspection, the ideal result of $F\rightarrow1$
can be achieved only if $\frac{\sigma_{n}}{\bar{n}}\rightarrow0$,
and either $\epsilon\rightarrow1$ or $\bar{n}\rightarrow\infty$.

We now compute $\bar{n}$ and $\sigma_{n}^{2}$ for a realistic optical
cycling process. We define the branching fraction to dark states,
which are lost from the optical cycle, to be $b_{\ell}$. We assume
that each particle interacts with the excitation laser light for a
time $T$, during which the scattering rate of a particle in the optical
cycle is $r$. Therefore, an average of $rT$ photons would be scattered
in the absence of decay to dark states, i.e. when $b_{\ell}=0$. (All
of our results hold for a time-dependent scattering rate $r(t)$,
with the substitution $rT\rightarrow\int r(t)dt$.) Note that in the
limit $rT\rightarrow\infty$, $1/b_{\ell}$ photons are scattered
per particle on average. Recall that the number of photons scattered
from the $i$-th particle, when projected to a given state, is $n_{i}$.
We define the probability that a particle emits exactly $n_{i}$ photons
to be $P(n_{i};\,rT,b_{\ell})$. This probability distribution can
be computed by first ignoring the decay to dark states. For the case
where $b_{\ell}=0$, the number of photons emitted in time $T$ follows
a Poisson distribution with average number of scattered photons $rT$.
For the more general case where $b_{\ell}>0$, we assign a binary
label to each photon depending on whether it is associated with a
decay to a dark state. Each decay is characterized by a Bernoulli
process, and we use the conventional labels of ``successful'' (corresponding
to decay to an optical cycling state) and ``unsuccessful'' (corresponding
to decay to a dark state) for each outcome. Then $P(n_{i};\,rT,b_{\ell})$
is the probability that there are exactly $n_{i}$ events in the Poisson
process, all of which are successful, or there are at least $n_{i}$
events such that the first $n_{i}-1$ are successful and the $n_{i}$-th
is unsuccessful. (For concreteness, we have assumed that \textquotedblleft unsuccessful\textquotedblright{}
decays, i.e., those that populate dark states, emit photons with the
same detection probability as all successful decays. The opposite
case, in which decays to dark states are always undetected, can be
worked out with the same approach and leads to similar conclusions.)
Direct calculation gives
\[
\bar{n}=\frac{1-e^{-b_{\ell}rT}}{b_{\ell}}{\rm \,and}
\]

\[
\sigma_{n}^{2}=\frac{1-b_{\ell}+e^{-b_{\ell}rT}b_{\ell}(2b_{\ell}rT-2rT+1)-e^{-2b_{\ell}rT}}{b_{\ell}^{2}}.
\]

\begin{figure}
\includegraphics[width=8cm]{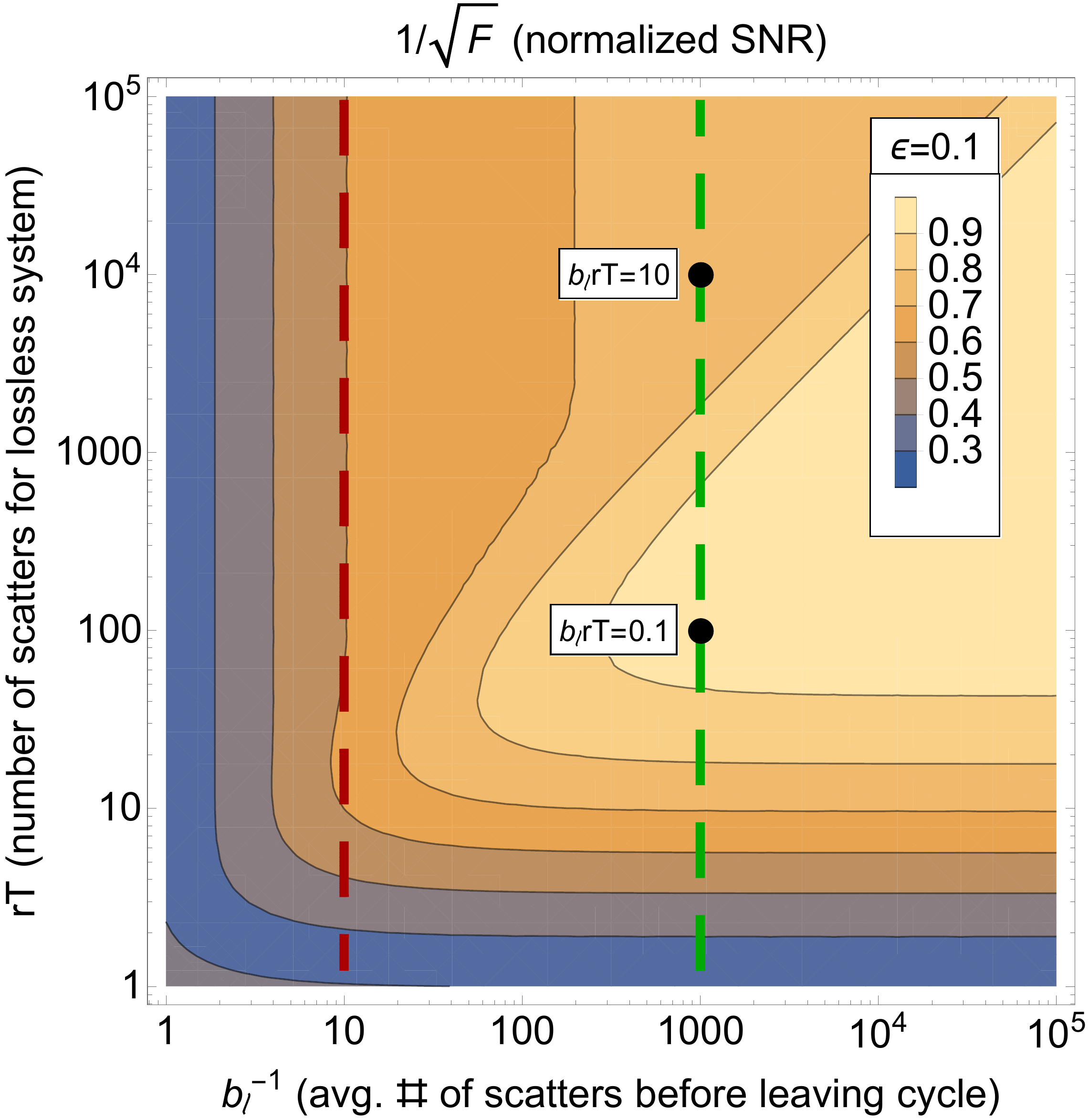}\caption{$1/\sqrt{F}$, the SNR resulting from Eq. \ref{eq:sigmaPhi}, normalized
to the ideal case of the QP limit ($F=1$). This plot assumes $\epsilon=0.1$.
When few photons per particle can be detected, i.e., when $\epsilon/b_{\ell}\ll1$
(far left of plot), cycling to very deep completion $(b_{\ell}rT\gg1$)
does not significantly affect the SNR. Even when one photon per particle
can be detected on average, i.e., when $\epsilon/b_{\ell}=1$ (dashed
red line), the SNR never exceeds roughly half its ideal value. By
further closing the optical cycle, i.e. such that $\epsilon/b_{\ell}\gg1$
(right of dashed red line), the SNR can be improved to near the optimal
value given by the QP limit. However, to reach this optimal regime,
the number of photons that would be scattered in the absence of dark
states, $rT$, must be small compared to the average number that can
be scattered before a particle exits the optical cycle, $1/b_{\ell}$.
For example, with $1/b_{\ell}=1,000$ (green dashed line) and $rT=100$
so that $b_{\ell}rT=0.1$ (lower circle), the SNR is more than 30\%
larger than in the case when $rT=10,000$ and $b_{\ell}rT=10$ (upper
circle). \label{fig:snr}}
\end{figure}

Therefore,

\begin{widetext}

\begin{equation}
F=1+\frac{1}{1-e^{-b_{\ell}rT}}\left(\frac{b_{\ell}}{\epsilon}+\frac{1-2b_{\ell}+2b_{\ell}e^{-b_{\ell}rT}(1-rT(1-b_{\ell}))-e^{-2b_{\ell}rT}}{1-e^{-b_{\ell}rT}}\right).\label{eq:sigmaPhi}
\end{equation}

\end{widetext}$\noindent$The behavior of the SNR (proportional to
$1/\sqrt{F}$) arising from Eq. \ref{eq:sigmaPhi} is illustrated
in Fig. \ref{fig:snr}.

To understand the implications of this result, we consider several
special cases, summarized in Table \ref{tab:special-cases}. We first
consider the simple case when cycling is allowed to proceed until
all particles decay to dark states, i.e., $b_{\ell}rT\rightarrow\infty$.
We refer to this as the case of ``cycling to completion.'' In this
case, for the generically applicable regime $\epsilon\leq\frac{1}{2}$
we find $F\geq2$, even as the transition becomes perfectly closed
($b_{\ell}\rightarrow0$). We can understand this result intuitively
as follows. As the optical cycling proceeds, the number of particles
that will still be in the optical cycle after each photon scatter
is proportional to the number of particles that are currently in the
optical cycle, $\frac{dP}{dn_{i}}\propto P$. Hence, we expect $P(n_{i};\,rT\rightarrow\infty,b_{\ell})\propto e^{-\alpha n_{i}}$
for some characteristic constant $\alpha$. In fact, one can show
that for $rT\rightarrow\infty$, this result holds with $\alpha\approx b_{\ell}$.
The width $\sigma_{n}$ of this exponential distribution is given
by the mean $\bar{n}$; that is, $\sigma_{n}\approx\bar{n}$. Therefore,
we should expect that cycling to completion reduces the SNR by a factor
of $\sqrt{F}=\sqrt{1+(\sigma_{n}/\bar{n})^{2}}\rightarrow\sqrt{2}$
compared to the ideal case of $F=1$, which requires $\frac{\sigma_{n}}{\bar{n}}=0$.

Surprisingly, this reduction in SNR can be partially recovered for
an imperfectly closed optical cycle, by choosing a finite cycling
time, $rT<\infty$, to minimize $\sigma_{\tilde{\phi}}$. The best
limiting case, as found from Eq. \ref{eq:sigmaPhi}, preserves the
condition that many photons are detected per particle, $rT\epsilon\gg1$,
but additionally requires that the probability of decaying to a dark
state remains small, $rTb_{\ell}\ll1$. In this case, photon emission
is approximately a Poisson process for which $\left(\frac{\sigma_{n}}{\bar{n}}\right)^{2}\approx\frac{1}{rT}\ll1$,
and the excess noise factor, $F$, does not have a significant contribution
from the variation in scattered photon number. The optimal value of
$rT$ for a finite proportion of decays to dark states, $b_{\ell}$,
and detection efficiency, $\epsilon$, lies in the intermediate regime
and can be computed numerically.

A special case of ``cycling to completion,'' which must be considered
separately, occurs when every particle scatters exactly one photon,
corresponding to parameter values $b_{\ell}=1$ and $rT\gg1$ so that
$\bar{n}=1$ and $\sigma_{n}=0$. As we have already seen, in this
case there is no contribution to the excess noise arising from variation
in the scattered photon number, and hence the SNR is limited only
by photon shot noise: $F=\frac{1}{\epsilon}$.

In atomic physics experiments with essentially completely closed optical
cycles, $b_{\ell}\approx0$, the limit $b_{\ell}rT\rightarrow\infty$
is not obtained even for very long cycling times where $rT\gg1$.
Instead, in this case $b_{\ell}rT\rightarrow0$ and hence $F\rightarrow1+\frac{1}{rT\epsilon}$,
which approaches unity as the probability to detect a photon from
each particle becomes large, $rT\epsilon\gg1$. Therefore, the reduction
in the SNR associated with the distribution of scattered photons does
not occur in this limit of a completely closed optical cycle.

\begin{table}
\centering{}%
\begin{tabular}{|c|c|c|c|}
\hline 
 & Condition & Sub-conditon & $F$\tabularnewline
\hline 
\hline 
1a & $b_{\ell}rT\rightarrow\infty$ &  & $2+b_{\ell}(\frac{1}{\epsilon}-2)$\tabularnewline
\hline 
1b & $b_{\ell}rT\rightarrow\infty$ & $\epsilon\leq0.5$ & $\geq2$\tabularnewline
\hline 
2a & $b_{\ell}rT\rightarrow0$ &  & $1+\frac{1}{rT\epsilon}+\frac{1}{2}b_{\ell}(\frac{1}{\epsilon}-2)$\tabularnewline
\hline 
2b & $b_{\ell}rT\rightarrow0$ & $\epsilon rT\rightarrow\infty$ & 1\tabularnewline
\hline 
3a & \textbf{$b_{\ell}\rightarrow1$} &  & $\frac{1}{\epsilon}\frac{1}{1-e^{-rT}}$\tabularnewline
\hline 
3b & $b_{\ell}\rightarrow1$ & $rT\rightarrow\infty$ & $\frac{1}{\epsilon}$\tabularnewline
\hline 
\end{tabular}\caption{The excess noise factor $F$ in some special cases. (1a) All particles
are lost to dark states during cycling. (1b) With all particles lost
and realistic detection efficiency, $\epsilon\leq0.5$, $F\geq2$.
(2a) No particles are lost to dark states. (2b) No particles are lost,
but many photons per particle are detected. The QP limit is reached.
(3a) Up to one photon can be scattered per particle. (3b) Exactly
one photon is scattered per particle and the photon shot noise limit
is reached.\label{tab:special-cases}}
\end{table}

We have also considered how the additional noise due to optical cycling
combines with other noise sources in the detection process. For example,
consider intrinsic noise in the photodetector itself. Commonly, a
photodetector (such as a photomultiplier or avalanche photodiode)
has average intrinsic gain $\bar{G}$ and variance in the gain $\sigma_{G}^{2}$,
with resulting excess noise factor $f=1+\frac{\sigma_{G}^{2}}{\bar{G}^{2}}$.
Including this imperfection in the model considered here leaves Eq.
\ref{eq:sigmaPhi} unchanged up to the substitution $\epsilon\rightarrow\epsilon/f$.
Similar derivations can be performed assuming a statistical distribution
of $N$ or $\phi$ to obtain qualitatively similar but more cumbersome
results.

In conclusion, we have shown that a quantum phase measurement, with
detection via laser-induced fluorescence using optical cycling on
an open transition, when driven to completion, incurs a reduction
in the SNR by a factor of $\sqrt{2}$ compared to the QP limit when
the optical cycle is driven to completion. This effect has been understood
as due to the distribution of the number of scattered photons for
this particular case. This reduction of the SNR does not occur for
typical atomic systems, where decay out of the optical cycle and into
dark states is negligible over the timescale of the measurement. An
expression for the SNR has been derived for the general case, in which
the cycling time is finite and the probability of decay to dark states
is non-zero. For a given decay rate to dark states, an optimal combination
of cycling rate and time can be computed numerically to obtain a SNR
that most closely approaches the QP limit. This ideal limit can be
obtained only when the photon cycling proceeds long enough for many
photons from each atom or molecule to be detected, but not long enough
for most atoms or molecules to exit the optical cycle by decaying
to an unaddressed dark state.
\begin{acknowledgments}
This work was supported by the NSF.
\end{acknowledgments}

\bibliographystyle{apsrev4-1}
\bibliography{cyclingPaperBibFinal}

\end{document}